\documentclass[conference,letterpaper]{IEEEtran}
\usepackage[utf8]{inputenc} 
\usepackage[T1]{fontenc}
\usepackage[cmex10]{amsmath}
\usepackage{amssymb, amsthm}
\usepackage{dsfont}
\usepackage{array}
\usepackage{graphicx}
\usepackage{epsfig, epstopdf}
\usepackage{cite}
\usepackage{algorithm, algorithmic}
\usepackage{color}
\usepackage{cite}

\newtheorem{rem}{Remark}

\newtheorem{prop}{Proposition}
\newtheorem{definition}{Definition}

\newtheorem{cor}{Corollary}

\newcommand\ceil[1]{\lceil#1\rceil}

\begin{document}

\title{Code Design Principles for Ultra-Reliable Random Access with Preassigned Patterns}

\author{%
	\IEEEauthorblockN{Christopher Boyd, Roope Vehkalahti and Olav Tirkkonen}
	\IEEEauthorblockA{Department of Communications and Networking\\
		Aalto University, Finland\\
		Email: \{christopher.boyd,roope.vehkalahti,olav.tirkkonen\}@aalto.fi}
	\and
	\IEEEauthorblockN{Antti Laaksonen}
	\IEEEauthorblockA{Department of Computer Science\\
		University of Helsinki, Finland\\
		Email: ahslaaks@cs.helsinki.fi}
}

\maketitle

\begin{abstract}
	

We study medium access control layer random access under the assumption that the receiver can perform successive interference cancellation, without feedback. During recent years, a number of protocols with impressive error performance have been suggested for this channel model. However, the random nature of these protocols causes an error floor which limits their usability when targeting ultra-reliable communications. In very recent works by Paolini et al. and Boyd et. al., it was shown that if each user employs predetermined combinatorial access patterns, this error floor disappears. In this paper, we develop code design criteria for deterministic random access protocols in the ultra-reliability region, and build codes based on these principles. The suggested design methods are supported by simulations.

\end{abstract}



\section{Introduction}

We consider contention-based multiple access in a situation where ultra-reliable low latency communication (URLLC) is required. Contention between accessing users makes achieving ultra-reliability particularly challenging. Guaranteeing grant-free access for uncoordinated machine-type users with near certainty necessitates the reconsideration of classical methods from the position of optimising for errors rather than throughput. This entails the design of new coding schemes, exploration of the trade-offs involved, and the acceptance of strict limitations. 

Modern variations on classical random access protocols have been recently introduced, including contention resolution diversity slotted ALOHA (CRDSA)~\cite{ICALOHA} and irregular repetition slotted ALOHA (IRSA)~\cite{Liva}. As in classical diversity slotted ALOHA (DSA)~\cite{ALOHA,DSA}, these medium access control (MAC) layer techniques consider synchronous communication on the random access channel (RACH) organised into frames of sequential time slots, with users transmitting their access packets along with a number of repetitions inside a frame. The protocols exploit successive interference cancellation (SIC) with repetition coding diversity to improve RACH throughput. 

In these protocols, the positions and/or the number of access packet replicas in a frame are selected at random, effectively allowing them to support an infinitely large user population $N$. However, this randomness guarantees an irreducible error floor caused by a non-zero probability of collisions, which jeopardises ultra-reliability. In~\cite{Paolini,IJWIN} it was shown that a deterministic code can yield repetition patterns that reduce this error floor significantly, while maintaining a reasonably large (but fixed) population. It was suggested that these codes be designed such that, as long as the number of simultaneously active users is below a given threshold $M<<N$, decoding will always succeed. We denote this property as $M$-interference cancelling or $M$-IC. In~\cite{Paolini}, this criterion was found through a link between LDPC parity-check matrices and random access code design, originally presented in~\cite{Liva}, while in \cite{IJWIN} it was discovered through combinatorial reasoning.

In this paper, we design deterministic random access codes for the ultra-reliability region, targeting a packet loss rate less than $10^{-5}$. We begin by considering simple deterministic variations on CRDSA (D-CRDSA), where the number of repetitions is fixed, and compare them to their original random versions. Such deterministic codes are always $2$-IC, and were considered in~\cite{IJWIN} with repetition factor $3$. Here, we show that larger code weights might maximise the user population $N$ and minimise the packet loss rate. 

More complex coding strategies are then considered, that can guarantee $M$-IC for $M>2$. We first show how the classical LDPC design principles used in \cite{Paolini}, avoiding small loops in the graph corresponding to the parity check matrix, leads to RACH codes that can only support small user populations. This result underlines that, while LDPC and RACH code design problems are to some extent equivalent~\cite{Liva}, the rate, block length and error probability regions of interest make some of the classical LDPC criteria less effective, or even counter-productive.

We then study how the limitations of LDPC based code design for URLLC can be overcome by using a combinatorial approach. In \cite{IJWIN}, we suggested the use of \emph{superimposed codes}~\cite{Kautz} as RACH codes that guarantee $M$-IC for $M>2$. A large class of such codes can be build from \emph{Steiner systems}. Here, we study Steiner systems as URLLC RACH codes and discuss how their usage differs from how they are classically used in LDPC code design. We also point out the limitations of any approach that is based on superimposed codes.


Finally, we study the trade-off between guaranteed decodability threshold $M$ and the supportable user population $N$, in the case where $M=3$. 

\section{System Model}

Consider a fixed population of $N$ machine-type users, who randomly access the channel resources and transmit an information packet to a central receiver in an uncoordinated and grant-free manner. This occurs during synchronized MAC frames of $n<<N$ time slots, each accommodating a single access packet. Users sporadically generate a packet and become active, attempting to transmit it during the next frame interval, along with $k<n$ repetitions arranged according to a codeword or pattern. It is assumed that these patterns may be preallocated to the users, and that each access packet contains pointers to the locations of its replicas. The collection of access patterns can thus be understood as a code consisting of $N$ binary vectors of length $n$, where 1's correspond to slots containing a copy of the transmitted packet.


In this paper, we consider a collision channel model, along with a receiver capable of performing perfect successive interference cancellation. This involves the receiver observing an entire frame; decoding any available interference-free packet; obtaining the location of its replicas; removing that users contribution to the frame; and iterating this process as many times as possible. We are interested in $M$-IC access codes, which means that any set of at most $M$ distinct codewords can be successfully recovered by the interference cancelling receiver. 
Ideally, these codes would be designed such that a large population of users may operate in an uncoordinated fashion with some certainty of limited interference. As such, we concentrate on the low activity scenario, where it is probable that the number of active users during a frame is $\leq M$. We analyse the decoding performance of the SIC receiver under Poisson arrivals, where the expected number of users active in a slot is given by the access intensity $\lambda$. The codes are compared with respect to the packet error rate, i.e. the fraction of the total number of transmitted packets that could not be decoded.



\section{ CRDSA vs D-CRDSA: the impact of pattern weights}\label{RandvsDet}

CRDSA is a RACH scheme wherein each active user randomly places some fixed number of repetitions of their access packet into the frame. A surprising result, observed for example in~\cite{Waterfall}, is that such a simple strategy can sometimes outperform more complicated irregular coding schemes in the ultra-reliability region. However, such schemes still suffer from an error floor. The limiting factor in the performance of such codes in the ultra-reliability region is a result of the \emph{birthday paradox}, i.e. a higher than intuitively expected probability that at least two active users simultaneously generate the same pattern~\cite{IJWIN}.  

In \cite{IJWIN}, we suggested a simple way of avoiding this error floor by using deterministic codes (D-CRDSA), where each user is uniquely allocated a weight $k$ access pattern. Such codes always satisfies the $2$-IC condition and can support at maximum of $n\choose k$ users. From this perspective, CRDSA can be viewed as a method where the active users are randomly choosing their access patterns from the selection of $n$ choose $k$ patterns, with the possibility that two or more users choose the same pattern. In the paper, the error performance of D-CRDSA and CRDSA with repetition factor $k=3$ was compared, and the superior performance of the deterministic approach was confirmed. However, there is no reason why $k=3$ should be the optimal number of repetitions. Additionally, it is not obvious if the gains observed in~\cite{IJWIN} would remain considerable when the number of repetitions in D-CRDSA and CRDSA is optimised. Intuitively, if the weight of the patterns $k$ is increased, then $n\choose k$ will also increase, and the effects of the birthday paradox can be expected to diminish along with the gains for D-CRDSA.
	
\begin{figure}[t]
	\centering
	\includegraphics[scale=0.57]{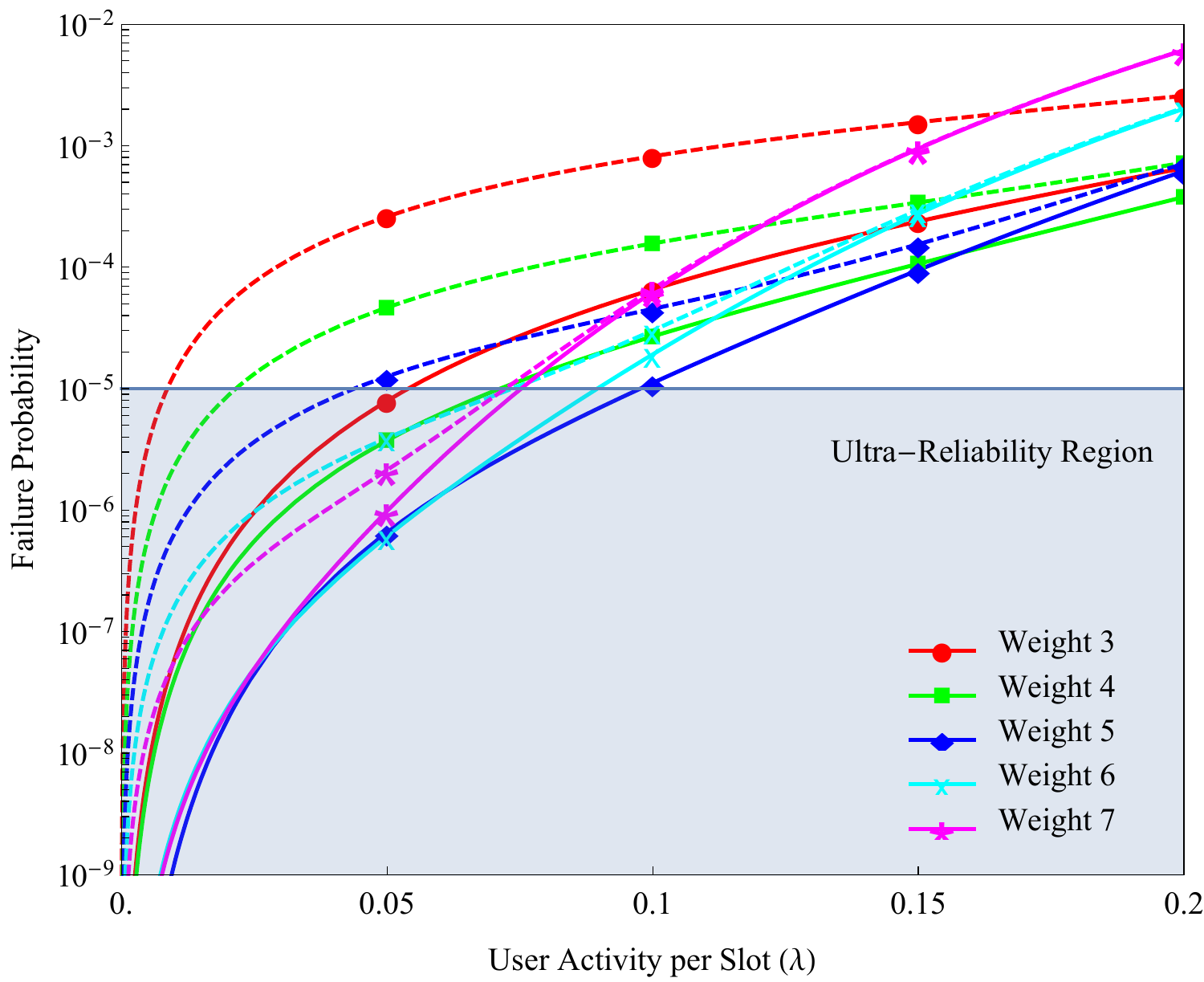}
	\caption{Simulated performance of D-CRDSA (solid) and CRDSA (dashed) in frames of 24 timeslots, under Poisson arrivals, for various uniform weights.}
	\label{fig:weights}	
	\vspace{-5mm}
\end{figure}

Figure~\ref{fig:weights} compares the packet decoding failure probabilities of deterministic and random codes of different weights, in access frame of $24$ timeslots. It can be observed that D-CRDSA outperforms the corresponding random codes in the ultra-reliability region, as expected. Also of note is that increasing the number of repetitions in D-CRDSA to $5$ improves the packet error rate, but beyond that the performance  deteriorates. On the other hand, the error performance of CRDSA continues to improve with increasing repetition factor. This can be contributed to the fact that, when $k=6$ and ${24\choose 6}=134596$, the effect of birthday paradox begins to fade.  
However, as the error performance of D-CRDSA is always better or equivalent to that of the corresponding CRDSA~\cite{IJWIN}, we can not improve beyond the performance of D-CRDSA with $7$ repetitions in the ultra-reliability region. 

Both the error probability region and the frame size affect the optimal number of repetitions for D-CRDSA. As far as we know, there does not yet exist a proper theory for predicting this number. Finally, we note that when using D-CRDSA, in selecting the optimal number of repetitions from an error probability point of view, we also implicitly decide the number of supportable users. In the above example, the optimal $5$ repetitions can support  ${24\choose 5}=42504$ users.

\section{Code Design for URLLC RACH}

In~\cite{IJWIN}, the most simple deterministic and uniform weight codes were discussed. Here, we consider more developed code designs and compare LDPC based and combinatorial approaches to the problem. Earlier, we saw how the weight of the patterns impacted the performance of RACH codes. However, the crucial difference between D-CRDSA and CRDSA is not related to the weight, but the fact that D-CRDSA codes always satisfy the $2$-IC condition. As suggested in~\cite{Paolini,IJWIN}, an obvious next step is to find codes that guarantee $M$-IC for $M>2$.

In \cite{IJWIN}, we provided the following sufficient and necessary condition for a RACH code to be $M$-IC.
\begin{definition}[\hspace{-0.01mm}\cite{IJWIN}]\label{IC-code}
	Let us suppose we have a random access code $C$ with $N$ length $n$ binary patterns (vectors), and let us denote by $A$ the $n\times N$-binary matrix formed by these vectors. Then, if each sub matrix $Y$ formed of $M$ or less columns of $A$ has the property that at least one of its rows has weight one, then $C$ guarantees correct decoding as long as there is at most $M$ active users.
\end{definition}

\begin{rem}
This type of combinatorial structure has appeared in the literature elsewhere. In~\cite{tracing}, these codes are called $(\leq M, 1, n)$-\emph{locally thin codes}. As we will see, this definition can also be used to describe the stopping sets of a LDPC code. We note also related research concerning \emph{non-adaptive collision resolution algorithms} \cite{KG}. 
\end{rem}

\subsection{RACH codes from classical LDPC designs}

It is known that the design of good random access codes and parity-check matrices for LDPC codes are equivalent problems~\cite{Paolini}. Matrix $A$ in Lemma~\ref{IC-code} can be viewed as a parity-check matrix of a LDPC code in a binary erasure channel, and the performance of the LDPC code under iterative decoding is equivalent to that of the corresponding RACH code with a SIC receiver~\cite{Liva}. In LDPC literature it is typical to present matrix $A$ as a bi-partiate graph. Using this language, classical LDPC code design suggests that a good code should not have small stopping sets. This condition was also used for RACH code design in ~\cite{Paolini}. In the language of Definition~\ref{IC-code}, if a set of columns of matrix $A$ does not satisfy the condition given, these vectors form a stopping set. The $M$-IC condition is met when the smallest stopping set is of size $M+1$.

A universal design criterion used to build LDPC codes that limit small stopping sets is to construct parity-check matrices that avoid small loops in the corresponding bi-partiate graph~\cite{Urbanke}. This criterion is based on a result in~\cite{Neuwirth} where it was proven that if the graph does not have small loops, then neither can it have small stopping sets. It was shown that if the parity-check matrix of an LDPC code has  columns of weight at least $k$, and the corresponding graph is $4$-loop free, then the smallest stopping set has at least $k+1$ elements. For our purposes, this translates to the corresponding RACH code being at least $k$-IC. However, as the following result shows, the approach of eliminating small loops places strict limitations on the number of possible users for small frames.

Consider access frames of size $n$ and random access code with codewords having weight at least $k>1$ and supporting $N$ users, whose corresponding graph does not contain $4$-loops. From~\cite[Prop. 2.1]{Neuwirth} we have that $N <{n\choose 2}$,
which implies that eliminating all loops of size $4$ or smaller significantly reduces the number of possible users. For example, when $n=24$ the maximum supportable user population is only $276$. 
In order to support large numbers of users, a different method of guaranteeing the $M$-IC condition for $M>2$ is required.

The reason for this partial failure of classical LDPC code design methods in small frames is related to the unusual rate region. Consider the matrix $A$ in Definition \ref{IC-code}. The corresponding LDPC code would have rate at least $\frac{N-n}{N}$. For example, if the frame size is $n=24$ and weight $k=5$ D-CRDSA is used, the corresponding LDPC code would have rate at least $0.99943..$. This rate region is atypical for error correcting codes. For comparison, in \cite{Paolini} the case of $n=200$ and $N=2000$ was considered, leading to a rate region which is still reasonable for classical LDPC methods.


\subsection{RACH codes from superimposed codes}
 
Consider the $n\times N$ matrix whose columns are  weight $k$ binary codewords of length $n$. The no $4$-loops condition can be seen to be  equivalent to the condition that any pair of columns (or rows) of the matrix share a $1$ on at most one position. This last condition is called the row-column (RC) condition~\cite{GallagerRC} and, as mentioned previously, it guarantees that the code is $k$-IC. 

In \cite{IJWIN}, it was suggested that more general \emph{superimposed codes} that guarantee $M$-IC could be employed as RACH codes. Using the notation of Definition \ref{IC-code}, we have
\begin{definition}[\hspace{-0.01mm}\cite{Kautz}]\label{def:matrixform}
	The set of patterns $A$ is a $M$-superimposed code if any $n\times M$ matrix formed from columns of $A$ has a submatrix consisting of an $M\times M$ permutation matrix. 
\end{definition}


Let us now present some known constructions of superimposed codes and assess their performance as RACH codes. In order to do this, we first move to set theoretical notation. Consider a random access code $C$ of frame-size $n$. By forming an index set $S=\{1,\dots,n\}$ the code $C$ can be seen as a collection of (different) subsets $A_j$ of $S$, where a subset corresponding a binary vector consist of all the elements of $S$, where the vector has $1$. We will denote the corresponding collection of such subsets by $\mathcal{A}$. In this notation, superimposed codes can be defined in a set theoretic language.

\begin{definition}[\hspace{-0.01mm}\cite{erdos}]\label{def:spernerSys}
	A collection of sets $\mathcal{A}$ is \emph{$M$-covering free} if it satisfies the following condition. Given any subset $X\in \mathcal{A}$, there does not exist sets $Y_1, ..,Y_{M-1} \in \mathcal{A}/ {X}$ such that  
	$$
	X\subseteq  Y_1 \cup Y_2 \cup.... \cup  Y_{M-1},
	$$
	Where $Y_i \in \mathcal{A}$. 
\end{definition}


We now provide a few examples of known superimposed code constructions and contrast them to codes satisfying the RC-condition.

\begin{prop}[\hspace{-0.01mm}\cite{erdos}]\label{sufficient}
	Let us suppose we have a collection $\mathcal{A}$ of subsets $A_i$, each having at least $k$ elements and satisfying
	$$
	|A_i\cap A_j|\leq v,
	$$
	when $i\neq j$.
	Then we have that the RACH code corresponding to $\mathcal{A}$ is $\ceil{\frac{k}{v}}$-superimposed.
\end{prop}

When translated into the language of binary vectors, for $v=1$ we recover the RC-condition. For larger $v$, this condition guarantees that given two codewords, they can have 1's on at most $v$ common positions. When the weight $k$ of the codewords is large enough, this condition is sufficient for guaranteeing the $\ceil{\frac{k}{v}}$-IC property. However, we note that while this condition does guarantee that the corresponding graph does not have small stopping sets, it will typically have $4$-loops when $v>1$. This simple result enables the use of known combinatorial constructions in code design for the RACH.

\begin{definition}\label{steiner}
	A Steiner system with parameters $t, k, n$, written $S(t,k,n)$, is an $n$-element set $S$ together with a set of $k$-element subsets of S (called blocks) with the property that each $t$-element subset of $S$ is contained in exactly one block. 
\end{definition}

Consider a Steiner system $S(t,k,n)$ with blocks $A_i$ and $A_j$. We have that $|A_i\cap A_j|\leq t-1$, as otherwise $A_i$ and $A_j$ would have at least $t$ elements in common. Together with Proposition \ref{sufficient}, this suggests the following.

\begin{cor}[\hspace{-0.01mm}\cite{erdos}]\label{Steinercor}

Steiner system $S(t,k,n)$ is $\ceil{\frac{k}{t-1}}$-superimposed.
\end{cor}
Note that $2$-designs (with $t=2$) have been used in LDPC code design, as they guarantee avoiding $4$-loops in the corresponding graph \cite{LDPCdesign}. Designs with $t>3$ have also been used, but they have been modified to also meet the no $4$-loops condition.

\begin{figure}[t]	
	\centering
	\includegraphics[scale=0.57,trim={5cm 0 0 0},clip]{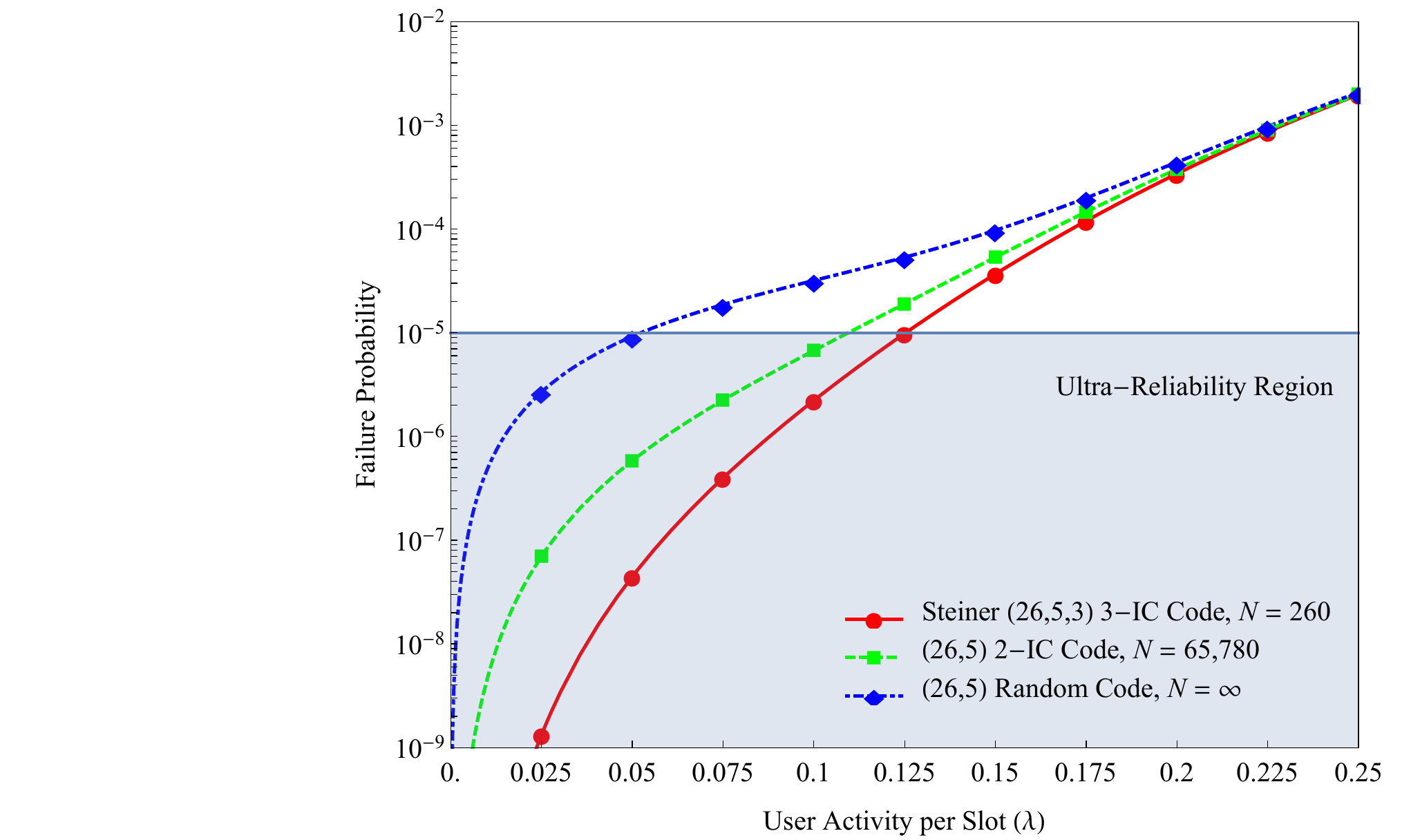}
	\caption{Simulated performance of the Steiner (3,5,26) system as a 3-IC random access code (solid) compared with the equivalent length 2-IC (dashed) and random (dot dashed) codes in frames of 26 timeslots, under Poisson arrivals.}
	\label{fig:2}
	\vspace{-5mm}
\end{figure}

The user population $N$ supported by $2$-designs is rather limited. For example, Steiner systems $S(2,3,27)$ and $S(2,3,67)$ produces $3$-IC codes that support only $N=117$ and $N=737$ users, respectively \cite{Steiner}. However, the more flexible code design principles we have presented here allow for the consideration of other systems of significantly larger size. For example, there exists a Steiner system $S(3,5,26)$ with $260$ elements and Steiner system $S(3,5,65)$ with $4368$ elements. According to Corollary~\ref{Steinercor}, these code are still $3$-IC, despite the fact that the corresponding graphs have very large number of $4$-loops. Figure~\ref{fig:2} illustrates how a $3$-IC random access code based on Steiner system $S(3,5,26)$ significantly outperforms D-CRDSA code of all weight $5$ codewords and the corresponding CRDSA code in terms of the error rate in the ultra-reliability region. While $N$ is limited to only $260$ users, it remains a reasonable user population for such small frames (26 timeslots). 

However, already for $M=3$ it becomes difficult to find large superimposed codes. For example, for frames of size $n=26$, the Steiner system with $260$ codewords is the largest we are aware of. A general asymptotic result~\cite{bounds} is that in order to always correctly decode $M$ out of $N$ possible users, the frames must be of size $n=\Omega\left(\frac{M^2\log(N)}{\log(M)}\right)$.

This shows that at most $\sqrt{n}$ of $N$ users may be active when the frame size is $n$. While the cost of additional active users is only in a logarithmic class, having even a modest number of active users forces the use of very large frames. This obviously restricts the usability of superimposed codes in the case of small frames.

We conclude by stating a general code design criterion for uniform weight RACH codes. While we discussed Steiner systems and D-CRDSA codes from the $M$-IC perspective, they also have other properties that make them beneficial as random access codes. These properties may be generalised into the following design principles for URLLC RACH codes. Consider once more the $n\times N$ matrix whose columns are the codewords of code $C$. We require that
\begin{itemize}
	\item[1.] Each column has weight $k$.
	\item[2.] Each row has weight $\gamma$.
	\item[3.] Given two columns, they have 1 in at most $v$ common positions.
\end{itemize}
This set of conditions is clearly a relaxation of the classical LDPC code design principle, where  $v=1$ in the third condition. 

\begin{rem}
Access patterns based on Steiner system $(3,5,26)$ were already used in~\cite{globecomm} by a subset of the authors of this paper, but without theoretical background. The simulations there were performed in a factory environment assuming fading channel conditions and $4$ access points.
\end{rem}


\section{Lower bound on the supportable user population for $3$-IC RACH codes}
The aforementioned RACH codes which satisfied the $3$-IC condition could only support relatively small user populations. 
Here, we consider the problem of finding $3$-IC RACH codes in size $n$ frames that support the maximum number of users $N$. Let $N(n)$ denote the size of such a code. For example, $N(3)=4$, because we can choose the patterns $\{100,010,001,111\}$ and it is not possible to construct a code with more patterns.

This can be seen as a combinatorial optimization problem where the maximum-size set of patterns of length $n$ that does not contain any \emph{forbidden triplet}, i.e., a subset of three patterns where no position has weight one, is sought. For example, $\{010,100,110\}$ is a forbidden triplet. Since the problem of finding a maximum-size subset with no forbidden subsets is equivalent to the NP-hard maximum set cover problem, an efficient algorithm is unlikely to exist and we restrict ourselves to $3$-IC codes instead of general $M$-IC codes.

\begin{table}[htbp]
  \renewcommand{\arraystretch}{1.3}
  \caption{Bounds for $3$-IC RACH codes.}
  \label{tab:bounds}
  \centering
  \begin{tabular}{r|r|r|r|r|r|r|r|r|r|r}
    
     $n$ &  1&2&3&4&5 &6 &7 &8        &9       &10 \\\hline
     $N(n)$&1&2&4&7&11&18&28& $\ge$ 44& $\ge$ 67&$\ge$ 102
    
    \end{tabular}
\end{table}

Table \ref{tab:bounds} includes bounds for $N(n)$ for $n=1,2,\dots,10$. The values for $n \le 7$ are exact, but only lower bounds are known for larger values of $n$. While the values for $n \le 6$ can be determined using a simple backtracking algorithm, the problem becomes much more difficult for larger values of $n$. To discover the codes for $n=7$ and $n=8$, we used a search algorithm that prunes the search tree by exploiting symmetries in codes. More precisely, we only considered codes where the set of patterns with weight one and two is lexicographically minimal when it is allowed to permute rows and columns.

We noticed that maximum-size codes for $n \le 7$ can also be generated using the following construction: To create a code for size $n+3$ frame, take a code with $n$ slots and for each pattern $x$ in that code, add patterns $100x$, $010x$ and $001x$ to the new code. In addition, add patterns $100z$, $010z$ and $001z$ where $z$ is a zero pattern of length $n$. Finally, add an additional set of patterns of the form $111a$ where $a$ is a pattern of length $n$. There are many ways to choose the last set of patterns; we can always at least include the pattern where $a$ is a zero pattern.  This construction is an extended version of the Busschbach construction \cite{Busschbach} (described in \cite{cohen}) that only adds patterns $100x$, $010x$ and $001x$ to the code. Our construction yields bounds $N(8) \ge 43$, $N(9) \ge 67$ and $N(10) \ge 102$. However, for $n=8$, we found a larger code with $44$ patterns using our search algorithm.

Given a code for size  $n$ frame  and $N$ users, our construction creates a code for size $n+3$ frame  and $3(N+1)+t$ users where $t$ is the number of patterns in the last set. Since $t$ is always at least one, the construction shows that $N(n+3) \ge 3(N(n)+1)+1$. This is a small improvement over the earlier Busschbach construction which shows that $N(n+3) \ge 3 N(n)$. In practice, $t$ can be larger than one. For example, to create the codes with sizes $67$ and $102$, we could add $10$ and $15$ additional patterns, respectively. However, we do not know how to determine the maximum value of $t$.

As far as we know, at least for small frame sizes these codes are the largest known. However, the suggested codes are not always good RACH codes as the weight distributions of the codes are strongly fluctuating. A natural next step would be to modify (or prune) these codes so that the weights of the corresponding codewords would be roughly equal.



\vspace{1mm}
\section*{Acknowledgements}

This work was funded in part by the Academy of Finland (grant 299916) and EIT ICT (HII:ACTIVE).


\vspace{1mm}
\bibliographystyle{IEEEtran}

\end{document}